%% file: elasticrmi.tex
\documentclass[runningheads,a4paper]{llncs}

\usepackage{url}
\usepackage{hyperref}
\usepackage{listings}
\usepackage{amssymb}
\usepackage{amsfonts}
\usepackage{amsmath}
\usepackage{graphicx}
\usepackage{subfig}
\usepackage{distribalgo}
\usepackage{algorithm}
\usepackage{algorithmic}
\usepackage{verbatim}
\usepackage{wrapfig}
\usepackage{fixme}

\usepackage{listings}
\usepackage{color}

\definecolor{javared}{rgb}{0.6,0,0} 
\definecolor{javagreen}{rgb}{0.25,0.5,0.35} 
\definecolor{javapurple}{rgb}{0.5,0,0.35} 
\definecolor{javadocblue}{rgb}{0.25,0.35,0.75} 

\definecolor{dkgreen}{rgb}{0,0.6,0}
\definecolor{gray}{rgb}{0.5,0.5,0.5}
\definecolor{mauve}{rgb}{0.58,0,0.82}
 
\newcommand{\elasticobjects}{\textsc{ElasticRMI}}
\newcommand{\pool}{elastic object pool}

\renewcommand{\subparagraph}{}

\begin{document}

\title{Elastic Remote Methods}

\author{K. R. Jayaram\thanks{Thanks to Patrick Eugster and Hans Boehm for helpful feedback.}}
\institute{HP Labs, Palo Alto, CA.\email{jayaramkr@hp.com}}

\toctitle{Elastic Remote Methods}
\tocauthor{K. R. Jayaram}

\maketitle

\input{abstract}

\input{introduction}

\input{overview}

\input{details}

\input{evaluation}

\input{relatedwork}

\input{conclusion}

\bibliographystyle{plain}

\bibliography{elasticobjects}

\end{document}

%% file: abstract.tex
\begin{abstract}
For distributed applications to take
full advantage of cloud computing systems, we need middleware systems
 that allow developers to build elasticity management components
right into the applications.

This paper describes the design and implementation of ElasticRMI, a
middleware system that (1) enables application
developers to dynamically  change the number of (server) objects available to
handle remote method invocations with respect to the application's workload,
without requiring major changes to clients (invokers) of remote methods, (2)
enables flexible elastic scaling by allowing developers to use a combination
of resource utilization metrics and fine-grained application-specific
information like the properties of internal data structures to drive scaling
decisions, (3) provides a high-level programming framework that handles
elasticity at the level of classes and objects, masking low-level
platform specific tasks (like provisioning VM
images) from the developer, and (4) increases the portability of 
ElasticRMI applications across different private data
centers/IaaS clouds through Apache Mesos~\cite{mesos}.

\keywords{programmable elasticity, scalability, distributed objects}

\end{abstract}

%% file: introduction.tex
\section{Introduction}

\emph{Elasticity}, the key driver of cloud computing, is the ability of a distributed application to 
dynamically increase or decrease its use of computing resources, to 
preserve its performance in response to varying workloads. Elasticity can either be \emph{explicit} or \emph{implicit}. 

\noindent{\bf Implicit vs. Explicit Elasticity.} Implicit elasticity is typically
associated with a specific programming 
framework or a Platform-as-a-Service (PaaS) cloud. Examples of frameworks providing implicit elasticity in the 
domain of ``big data analytics'' are map-reduce (Hadoop~\cite{hadoop} 
and its PaaS counterpart Amazon Elastic Map Reduce~\cite{emr}), Apache Pig~\cite{pig}, Giraph~\cite{giraph}, 
etc. Implicit elasticity is handled
by the PaaS implementation and is not the responsibility of the programmer. Despite being unable
to support a wide variety of applications and computations, each of these systems
simplifies application
development and deployment, and employs distributed algorithms for elastic scaling that are 
optimized for its programming framework and application domain. 

Explicit Elasticity, on the other hand, is typically associated with 
Infrastructure-as-a-Service (IaaS) clouds and/or private data centers, which 
typically provide elasticity at the granularity
of virtualized compute nodes (e.g., Amazon EC2) or 
virtualized storage (e.g.,  Amazon Elastic Block Store (EBS)) in
a way that is agnostic of the application using these resources. It is the application developer's
or the system administrator's 
responsibility to implement robust mechanisms to monitor the application's
performance at runtime, request the addition or removal of resources, and perform
\emph{load-balancing}, i.e., redistribute the application's workload among the new set of
resources.

\noindent{\bf Programmable Elasticity.} To \emph{optimize} the performance of \emph{new or existing} distributed applications while
\emph{deploying or moving} them to the cloud, engineering robust elasticity management 
components is essential. This is especially vital for applications that do not fit the programming model
of (implicit) elastic frameworks like Hadoop, Pig, etc., but require
high performance (high throughput and low latency), scalability and elasticity -- the best example is the class of 
 \emph{datacenter infrastructure applications} like 
key-value stores (e.g., memcached~\cite{memcached}, Hyperdex~\cite{kvstore}), consensus protocols 
(e.g., Paxos~\cite{paxos}), distributed lock
managers (e.g., Chubby~\cite{chubby}) and message queues. Elasticity frameworks which
rely on externally observable resource utilization metrics (CPU, RAM, etc.) are
insufficient for such applications (as we demonstrate empirically in Section~\ref{sec:eval}). A distributed key value store, for example, may have high CPU utilization when there
is high contention to update a certain set of ``hot keys''. Relying on CPU utilization to 
simply add additional compute nodes will only degrade its performance further. Hence, there is an emerging need for a elasticity framework that bridges the gap between implicit and explicit elasticity, allowing the use \emph{fine grained} application specific 
metrics (e.g., size of a queue/heap, number of aborted transactions or average number of attempts to acquire certain locks) to build an elasticity management component right into
the application without compromising (1) security, by revealing application-level information to the 
cloud service provider, and (2) portability across different cloud vendors.

\noindent{\bf Why RMI?} Despite being criticized for introducing direct dependencies across nodes
through remote object references, Remote Method Invocation (RMI) and Remote Procedure Call (RPC) remain a popular 
paradigm~\cite{thrift} for distributed programming, because of their simplicity. Their popularity has led to the 
development of Apache Thrift~\cite{thrift} with support for RPC
across \emph{different} languages, and the design of cloud computing paradigms like RAMCloud~\cite{ramcloud}
with support for low-latency RPC. However, high-level support for elasticity is limited in existing RPC-like frameworks; and such support is vital to engineering efficient distributed applications and migrating existing applications to the cloud.

\noindent{\bf Contributions.} This paper
makes the following technical contributions:

\begin{enumerate}

\item \elasticobjects\ -- A framework for elastic distributed objects in Java:

\begin{enumerate}
 
 \item with the same simplicity and ease of use of the Java RMI, handling elasticity
 at the level of classes and objects, while supporting implicit and explicit elasticity. (Section~\ref{sec:overview})

\item that enables application developers to dynamically 
change the number of (server) objects available to handle remote method invocations
with respect to the application's workload, without requiring any change to clients (invokers) of remote
methods. (Section~\ref{sec:details})

\item enables flexible elastic scaling by allowing developers to use a combination of resource
utilization metrics and fine-grained application-specific information to drive scaling decisions. (Section~\ref{sec:details})

\end{enumerate}

\item A runtime system that handles all the low-level mechanics
of instantiating elastic objects, monitoring their workload, adding/removing additional
objects as necessary and load balancing among them. (Section~\ref{sec:runtime})

\item Performance evaluation of \elasticobjects\ using elasticity metrics
defined by SPEC~\cite{specmetrics} and using four \emph{existing} real world applications -- Marketcetera
financial order processing, Paxos consenus protocol, Hedwig publish/subscribe system and a distributed
coordination service. (Section~\ref{sec:eval})

\end{enumerate}

%% file: overview.tex
\section{\elasticobjects\ -- Overview}\label{sec:overview}

\begin{figure}[tb]
\begin{center}
\includegraphics[width=0.90\textwidth]{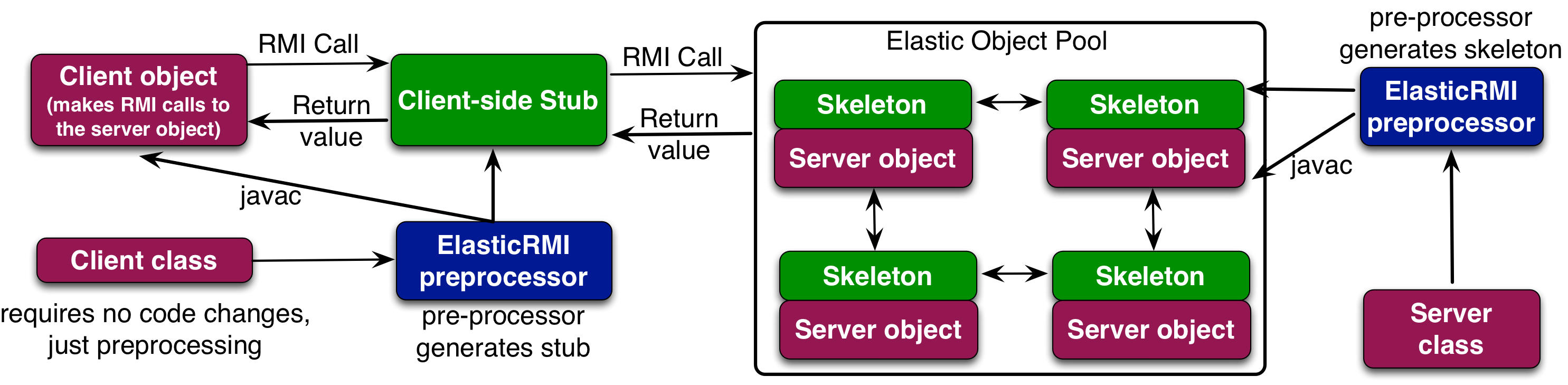}
\end{center}
\caption{\elasticobjects\ -- Overview}\label{fig:overview}
\end{figure}

Figure~\ref{fig:overview} illustrates the architecture of
\elasticobjects. In designing \elasticobjects, our goals are (1) to retain the simple
programming model of Java RMI, and mask the low level details 
of implementing elasticity like workload monitoring, load balancing,
and adding/removing objects from the application developer. (2) require
minimal changes, if any, to the clients of an object (3) make \elasticobjects\
applications as portable as possible across different IaaS cloud implementations.

\subsection{Elastic Classes and Object Pools}~\label{sec:classesintro}

An \elasticobjects\ application consists 
of multiple components (implemented as classes) interacting with 
each other. The basic elasticity abstraction in \elasticobjects\
is an elastic class. An elastic class is also a remote class, and 
(some) of its methods may be invoked remotely from another JVM. The 
key difference between an elastic and a regular remote class is that an 
elastic class is instantiated into a pool of 
objects (referred to as \pool), with each object executing on a separate JVM.
But, the presence of multiple objects in an elastic object pool is 
transparent to its clients, because the pool behaves as a single remote object.
Clients can only interact with the entire object pool, by invoking its remote 
methods. The interaction between a client and an elastic object pool is 
\emph{unicast} interaction, similar to Java RMI. The processing of the method 
invocation, i.e., the method execution happens at a single object in 
the elastic object pool chosen by the \elasticobjects\ runtime, and not the client.
 \elasticobjectsÕs runtime can redirect incoming method 
invocations to one of the objects in the pool, depending on various factors 
and performance metrics (Section~\ref{sec:runtime} describes load balancing in detail). 
The runtime automatically changes the size of the pool depending on the pool's
``workload''.
\elasticobjects\ is different from certain  
frameworks where the same 
method invocation is multicast and consequently executes on multiple (replicated) objects for fault-tolerance.

\subsection{Shared State and Consistency}

In Java RMI, the \emph{state} of a remote (server) object is a simple concept,
because it resides on a single JVM and there
is exactly one copy of all its fields. Clients of the remote object 
can set the value of instance fields by calling a remote method, and access
the values in subsequent method calls. In \elasticobjects, on the other
hand, the entire object pool should appear to the client as a single 
remote object -- thereby necessitating coordination between the objects in the
pool to consistently update the values of instance fields. For consistency, 
we employ an \emph{external} in-memory key-value store (HyperDex~\cite{kvstore} in our 
implementation) to store the state (i.e., \lstinline{public, private, protected} and 
\lstinline{static} fields) of the elastic remote object pool. The key-value store is \emph{not} used
to store local variables in methods/blocks of code and parameters in method 
declarations. Local variables are instantiated on the JVM in which the object resides. The key-value
store is shared between the objects in the pool, and executes on separate JVMs.

\subsection{Stubs and Skeletons}

\elasticobjects\ modifies the standard mechanisms used to implement RPCs and Java RMI,
as illustrated by Figure~\ref{fig:overview}.
The \elasticobjects\ \emph{pre-processor} analyzes elastic classes to generate
\emph{stubs} and \emph{skeletons} for client-server communication. As in Java RMI, 
a stub for a remote object acts as a client's local representative or proxy for 
the remote object. The caller invokes a method on \elasticobjects's local stub which 
(1) initiates a connection
with the remote JVM, serializes and marshals parameters, waits for the result of the
method invocation and unmarshals the return value/exception before returning it to
the sender, and (2) performs load balancing among the objects in the elastic pool as necessary.
The stub is generated by the \elasticobjects\ preprocessor and is different from the client application, to which the entire object pool appears as a single object, i.e., the existence of a
pool of objects is known to the stub but not to the client application. In the remote JVM, each object in the pool has a corresponding 
skeleton, which in addition to the duties performed in regular Java RMI, can also
perform dynamic load balancing based on the CPU utilization of the object and 
redirect all further method invocations to other objects in the pool after \elasticobjects\
decides to shut it down in response to decreasing workload.

\subsection{Instantiation of Object Pools in a Cluster}~\label{sec:stubsskeletons}

An elastic class can only be instantiated by providing a minimum and maximum number 
of objects that constitute its \pool. Obviously, instantiating all objects in the pool
on separate JVMs on the \emph{same} physical machine may degrade performance. Hence, \elasticobjects\
attempts to instantiate each object in a virtual node in a compute cluster. Virtual nodes 
can be obtained either (1) from IaaS clouds by provisioning and 
instantiating virtual machines, or (2) from a cluster management/resource sharing system like 
Apache Mesos~\cite{mesos}. Our implementation of \elasticobjects\ uses 
Apache Mesos~\cite{mesos} because it supports both clusters of physical nodes (in 
private data centers) or virtual nodes (from IaaS clouds). Mesos can also be viewed as a thin 
resource-sharing layer that manages a cluster of physical nodes/virtual machines. It
divides these nodes into ``slices'' (called \emph{resource offers} or slave nodes~\cite{mesos}), with 
each resource offer containing a configurable reservation of CPU power (e.g., 2 CPUs at 2GHz), 
memory (e.g., 2GB RAM), etc. on one of the nodes being managed. Mesos implements the ``slice'' 
abstraction by using Linux Containers (\url{http://lxc.sourceforge.net}) to 
implement lightweight virtualization, process isolation and resource guarantees (e.g., 2CPUs at 2GHz)~\cite{mesos}. While instantiating 
an elastic class, the \elasticobjects\ runtime requests Mesos for a specified 
number of slave nodes, instantiating an object on each slave node. 

Mesos aids in portability of \elasticobjects\ applications, just like the JVM aids
the portability of Java applications. Mesos can be installed on private data centers and 
many public cloud offerings (like Google Compute Engine, Amazon EC2, etc.). As long 
as Mesos is available, \elasticobjects\ applications can be executed, making them
portable across different cloud vendors.

\begin{figure}
\begin{center}
\includegraphics[width=0.8\columnwidth]{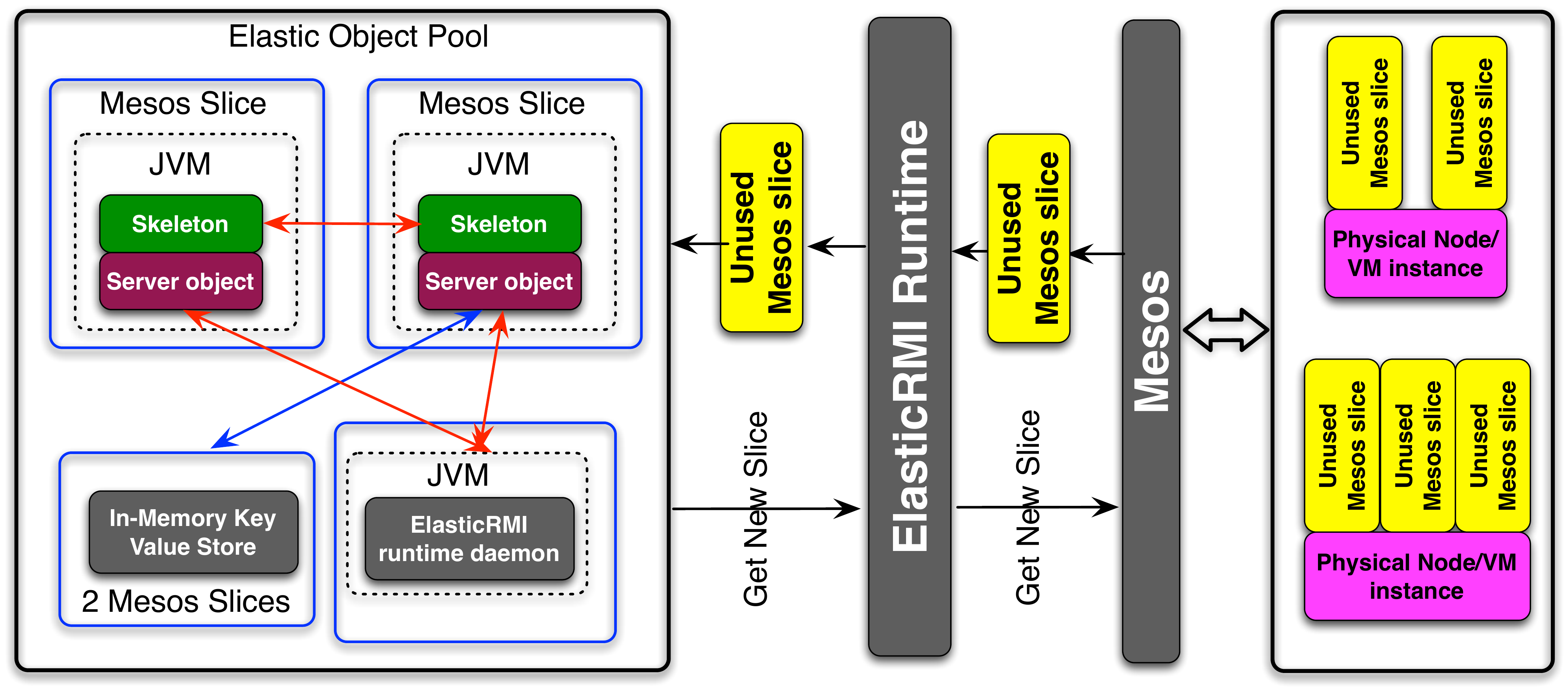}
\caption{\elasticobjects\ -- Server Side}\label{fig:serverside}
\end{center}
\end{figure} 

\subsection{Automatic Elastic Scaling}

The key objective of \elasticobjects\ is to change the number of objects in 
the elastic pool based on its workload. The ``workload'' of an elastic 
object can have several application-specific definitions, and consequently, 
\elasticobjects\ allows programmers to define the workload of an elastic 
class and specify the conditions under which objects should be added or 
removed from the elastic pool. This can be done by overriding select methods in the
\elasticobjects\ framework, as discussed in detail in (the next) Section~\ref{sec:programming}.
During the lifetime of the elastic object, the \elasticobjects\ runtime monitors a \pool's 
workload and decides whether to change its size either
based on default heuristics or by applying the programmer's logic by invoking the overridden methods
discussed above.

If a decision has been made to increase the size of the pool, the \elasticobjects\ runtime
interacts with the Mesos master node to request additional compute resources. If the request is
granted, \elasticobjects\ runtime instantiates the additional object, and adds it to the pool (See Figure~\ref{fig:serverside} for an illustration). 
If the decision is to remove an object, \elasticobjects\ communicates with its skeleton
to redirect subsequent remote method calls to other objects in the pool. Once redirection
starts, \elasticobjects\ sends a \textsc{shutdown} message to the object. The object
acknowledges the message, and waits for all pending remote method invocations to
finish execution or throw exceptions indicating abnormal termination. Then the object
notifies the \elasticobjects\ runtime that it is ready to be shutdown.  \elasticobjects\ terminates
the object and relinquishes its  slice to Mesos. This slice is then available to
other elastic objects in the cluster, or for subsequent use by the same elastic object if a  
decision is made in the future to increase the size of its pool.

%% file: details.tex
\section{Programming With \elasticobjects}\label{sec:details}

This section illustrates the use of \elasticobjects\ for both implicit and explicit elasticity
through examples, along with an overview of how to make such decisions with a \emph{global} view of
the entire application. Our implementation also includes a preprocessor similar to
\lstinline{rmic} which in addition to generating stubs and skeletons, converts \elasticobjects\ programs into plain
Java programs that can be compiled with the \lstinline{javac} compiler.

 \begin{figure}[tb]
\begin{lstlisting}[caption=]
interface Elastic extends Remote { } //marker interface used by preprocessor
//Optionally implemented by the application
abstract class Decider extends UnicastRemoteObject implements Elastic {
  abstract int getDesiredPoolSize(ElasticObject o);
}
abstract class ElasticObject extends UnicastRemoteObject {
  ElasticObject() //Default constructor
  ElasticObject(Decider d) //Consult d for scaling decisions
 
  void setMinPoolSize(int s) //Set min pool size
  void setMaxPoolSize(int s) //Set max pool size  
  void setBurstInterval(float ival) //Make scaling decisions every 'ival' ms
  void setCPUIncrThreshold(float t) //Add objects when CPU util > t
  void setCPUDecrThreshold(float t) //Remove objects when CPU until < t 
  void setRAMIncrThreshold(float t) //Add objects when RAM util > t
  void setRAMDecrThreshold(float t) //Remove objects when RAM until < t
     
  float getAvgCPUUsage() //Get CPU util averaged over burst interval
  public float getAvgRAMUsage() //Get RAM util averaged over burst interval
  int getPoolSize() //Get pool size
  //Returns average # of calls to each remote method over the burst interval
  HashMap<String,float> getMethodCallStats() {...}
    
  //Called by the runtime to poll each object about changes to the size of  
  //the pool. Can return positive or negative integers
  abstract int changePoolSize();
}
\end{lstlisting}
\caption{A snapshot of the \elasticobjects\ server-side API.}\label{fig:api}
\end{figure}

\begin{figure*}[htb]
\subfloat[Implicit elasticity\label{fig:implicitexample}]{
\begin{lstlisting}[linewidth=1.65in,caption=, frame=]
class CacheImplicit
  extends ElasticObject {
  CacheImplicit() {
    setMinPoolSize(5);   
    setMaxPoolSize(50);
  }
  ...
}
\end{lstlisting}}
\subfloat[Explicit elasticity using coarse-grained metrics.\label{fig:explicitexample}]{
\begin{lstlisting}[linewidth=3.4in,caption=,frame=l]
class CacheExplicit1 extends ElasticObject {
  CacheExplicit1() {
    setMinPoolSize(5); setMaxPoolSize(50);
    setBurstInterval(5*60*1000); //5mins
    setCPUIncrThreshold(85); setRAMIncrThreshold(70);
    setCPUDecrThreshold(50); setRAMDecrThreshold(40);
  } ...
}
\end{lstlisting}}
\caption{Example of two distributed cache classes implemented in \elasticobjects}
\end{figure*}

\subsection{\elasticobjects\ Class Hierarchy}\label{sec:programming}

A distributed application built using \elasticobjects\ consists of interfaces declaring methods
and classes implementing them. The key features of \elasticobjects\ API (Figure~\ref{fig:api}) are:

\begin{itemize}

\item \lstinline{java.elasticrmi} is the top-level package for programming \elasticobjects\
server classes.

\item An elastic interface is one that declares 
the set of methods that can be invoked from a remote JVM (client). All elastic interfaces 
must extend \elasticobjects's marker interface -- \lstinline{java.elasticrmi.Elastic}, 
either directly or indirectly.

\item The \lstinline{ElasticObject} class implements all the 
 basic functionalities of \elasticobjects. An application-defined class becomes elastic by implementing one or 
 more elastic interfaces and by extending \lstinline{ElasticObject}.

\end{itemize}

\subsection{Programming With Implicit Elasticity}

\elasticobjects\ supports implicit elastic scaling, using average CPU utilization 
across the objects in an elastic pool as the default coarse-grained metric. 
A time interval, referred to as the \emph{burst interval} (default 60s)
is used to decide whether to change the size of elastic object pool. 
\elasticobjects\ measures the average 
CPU utilization 
of each object in the elastic pool every 60s -- objects are added (in increments of 1 object) 
when the average utilization exceeds 90\% and removed when average utilization falls 
below 60\%. Figure~\ref{fig:implicitexample} shows an example of a distributed cache class (e.g., a web cache,
content/object cache) that relies on
\elasticobjects's implicit elasticity mechanisms. The programmer simply implements a cache
store based on some well-known algorithm and specifying the minimum and maximum size of the pool,
 without worrying about adapting to new resources and load balancing.

\subsection{Programming With Explicit Elasticity}

\elasticobjects\ also allows programmers to \emph{explicitly} define the workload of an elastic class 
and specify the conditions under which objects should be added or removed from the 
elastic pool. Workload definitions can either be coarse-grained or fine-grained. 

\paragraph{Coarse-grained Metrics.}

 A programmer can override the 
default burst interval, and the average CPU utilization thresholds used for 
changing the number of objects in the pool by calling the appropriate methods 
(\lstinline{setBurstInterval(...)}, \lstinline{setCPUIncrThreshold(...)} and 
\lstinline{setCPUDecrThreshold(...)}) 
in \lstinline{java.elasticrmi.ElasticObject} which is available in all elastic classes
since they extend \lstinline{ElasticObject} (see Figure~\ref{fig:api}).

Figure~\ref{fig:explicitexample} shows an example of a cache class 
that changes the CPU and memory (RAM) utilization thresholds that trigger
elastic scaling. The core \elasticobjects\ API includes 
specific methods to set CPU and memory thresholds because they are commonly
used for elastic scaling -- if both CPU and RAM thresholds are set,
the runtime interprets them using a logical OR, i.e., in the example
shown in Figure~\ref{fig:explicitexample}, the \elasticobjects\ runtime 
increases the size of the pool by in increments of 1 object every
five minutes, either if average CPU utilization exceeds 85\% or if 
average memory utilization exceeds 70\% across the JVMs in the 
\pool.

\paragraph{Fine-grained Metrics.}

\elasticobjects\ provides additional support, through the \lstinline{changePoolSize} method
(see Figure~\ref{fig:api}) which can be overridden by any elastic class. The runtime 
periodically (every ``burst interval'') invokes \lstinline{changePoolSize} to poll each object in the \pool,
about desired changes to the size of the pool. The method returns an integer -- positive or
negative corresponding to increasing or decreasing the pool's size. The values returned by the
various objects in the pool are averaged to determine the number of objects that have to be added/removed. The logic used
to decide on elastic scaling is left to the developer, and it may be based on 
(1) parameters of the JVM on which each object resides, (2) properties of 
shared instance fields of the elastic object, or of data structures used by the
object, e.g., number of pending client operations stored in a queue, and (3) metrics
computed by the object like average response time, throughput, etc. \elasticobjects\
allows classes to use only a single decision mechanism for elastic scaling, i.e., 
if \lstinline{changePoolSize} is overridden, then scaling based on CPU/Memory utilization
is disabled.

\begin{figure}[tb]
\begin{lstlisting}[caption=]
public class CacheExplicit2 extends ElasticObject {
  float avgLockAcqFailure, avgLockAcqLatency
  public int changePoolSize() {
    HashMap<String, float> sMap;
    sMap = getMethodCallStats();
        
    float putLatency = sMap.get("put").getLatency();
    float getLatency = sMap.get("get").getLatency();
    if(putLatency > 100 || putLatency > 3*getLatency) {
      if(avgLockAcqFailure > 50) return 0;
      if(avgLockAcqLatency >= 0.8*putLatency) return 0; else return 2;
    }
 }
}
\end{lstlisting}
\caption{A distributed cache class which relies on \elasticobjects's explicit elasticity support using fine-grained application-specific metrics.}\label{fig:finegrainedexample}
\end{figure}

Figure~\ref{fig:finegrainedexample} illustrates the use of \lstinline{changePoolSize}
to make scaling decisions. The \lstinline{CacheExplicit2} class is implemented
to use metrics specific to distributed object caches, e.g., \lstinline{avgLockAcqFailure}
(which measures the failure rate of acquiring write locks to ensure consistency during
a \lstinline{put} operation on the cache) and \lstinline{avgLockAcqLatency}
(which measures the average latency to acquire write locks) to make decisions about
changing the size of the \pool. In Figure~\ref{fig:finegrainedexample}, the 
\lstinline{CacheExplicit2} class does not add new objects to the pool
when there is a lot of contention. When the failure rate for
acquiring write locks (\lstinline{avgLockAcqFailure} is 
greater than 50\%) or when the predominant component of \lstinline{putLatency} 
is \lstinline{avgLockAcqLatency}, no additional objects are added to the pool
because there is already high contention among objects serving client requests 
to acquire write locks. When these conditions are \lstinline{false}, the size
of the pool is increased by two -- controlling the number of objects added is 
another feature of \lstinline{changePoolSize}.

\paragraph{Making Application-Level Scaling Decisions.}\label{sec:applscaling}

The mechanisms described above involve makes scaling decisions \emph{local to an elastic class},
and may not be optimal for 
applications using multiple elastic classes (where the application contains \emph{tiers} of elastic
pools). \elasticobjects\ also supports decision making at the level of the application using the 
\lstinline{Decider} class. It is the developer's responsibility
to ensure that elastic objects being monitored communicate with the 
monitoring components, either by using remote method invocations or through message 
passing. The \elasticobjects\ runtime assumes responsibility for calling
\lstinline{changePoolSize} method of the monitoring class to get the desired size of each 
\pool, and determines whether objects have to be added or removed. Due to space limitations,
we refer the reader to our tech report for additional details~\cite{techreport}.

\section{The \elasticobjects\ Runtime}\label{sec:runtime}

The runtime  
(1) handles shared state among the objects in an \pool, 
(2) instantiates each object of a pool, (3) performs load balancing, and (4) is responsible for fault-tolerance.

\subsection{Shared State and Consistency}


The objects in the \pool\ 
coordinate to update the state of its instance fields (\lstinline{public}, \lstinline{private},
\lstinline{protected}) and \lstinline{static} fields. For consistency,
we use HyperDex~\cite{kvstore}, a distributed in-memory key-value store (with strong consistency).
Using an in-memory store provides the same 
data durability guarantees as Java RMI which stores the state of instance fields in RAM in 
a single Java virtual machine heap. HyperDex is different from the distributed cache in
the examples of Section~\ref{sec:details}, which is used for illustration purposes only. The \elasticobjects\ preprocessor translates reads and writes of instance and static fields
into \lstinline{get(...)} and \lstinline{put(...)} method calls of HyperDex. 


\begin{wrapfigure}{r}{0.69\textwidth}
\vspace{-8mm}
\begin{lstlisting}[caption=]
class C1                 |class C1Processed {  
 extends ElasticObject { | void foo() {
  int x, z;              |  int x=Store.get("C1$x");
                         |  if(x==5) 
  void foo() {           |   Store.put("C1$z", 10);
    if(x = 5) z = 10;    | } 
  }                      | void bar() {  
  synchronized           |  while(!ERMI.lock("C1")) ;
    void bar() {         |  ...
    ...                  |  ERMI.unlock("C1"); 
  }                      | }
}                        |}
\end{lstlisting}
\caption{Handling shared state through an in-memory store (HyperDex~\cite{kvstore} here).}\label{fig:compiledexample}
\end{wrapfigure}

Figure~\ref{fig:compiledexample} shows a simple elastic class \lstinline{C1} and how it is transformed by the \elasticobjects\
preprocessor to insert calls to HyperDex (abstracted by \lstinline{Store}) and 
\elasticobjects's runtime (abstracted by \lstinline{ERMI}). For variable \lstinline{x},
\lstinline{C1$x} is used as the key in \lstinline{Store}. For synchronized
methods, \elasticobjects\ uses a lock per class named after the class -- the example
in Figure~\ref{fig:compiledexample} uses a lock called \lstinline{"C1"}.

If an elastic class has an instance or static field $f$, a method call on $f$ is handled as follows: \vspace{-4mm}

\begin{itemize}
\item If $f$ is a remote or elastic object, the method invocation is simply serialized and dispatched to the remote object or pool as the case may be.

\item \emph{Else}, if $m$ is \lstinline{synchronized}. the method call is handled as in Figure~\ref{fig:compiledexample}, but the runtime acquires a lock on $f$ through HyperDex. In this case, 
\elasticobjects\ guarantees mutual exclusion for the execution of $f.m(...)$ with respect to other methods of $f$.
  
\item \emph{Else}, (i.e., $m$ is neither remote, elastic nor synchronized), $f.m(...)$ involves retrieving $f$ from HyperDex and executing $m(...)$ locally
on an object in the \pool, and storing $f$ back into HyperDex after $m(...)$
has completed executing. 

\end{itemize}

\elasticobjects\ aims to increase parallelism and hence the
number of remote method executions per second when there is limited or no
shared state. Increasing shared state increases latency due to the 
network delays involved in accessing HyperDex. Having shared state and 
mutual exclusion through locks or synchronized methods further decreases parallelism.
However, we note that this is not a consequence of \elasticobjects, but rather
dependent on the needs of the distributed application. If the developer
manually implements all aspects of elasticity by using plan Java RMI and 
an existing tool like Amazon CloudWatch+Autoscaling, he still has to use
something like a key-value store to handle shared state. \elasticobjects\ cannot
and does not attempt to eliminate this problem -- it is up to the programmer
to reduce shared state. 
 
 Note that \elasticobjects\ does not guarantee a
transactional (ACID) execution of $m(...)$ with respect to other objects in the pool, and 
using \lstinline{synchronized} does not provide ACID guarantees \emph{either in RMI or in \elasticobjects}.

\subsection{Instantiation of Elastic Objects}
An elastic class can only be instantiated by providing a minimum ($\geq$ 2) and maximum number 
of objects that constitute its \pool\ (see Figure~\ref{fig:api}). During instantiation, 
if the minimum number of objects is $k$, ElasticRMI's runtime creates $k$ objects on $k$ new 
JVM instances on $k$ virtual nodes (Mesos slices), if $k$ virtual nodes are available from Mesos~\cite{mesos}. 
If only $l < k$ are available, then only $l$ objects are created. Under no circumstance does
\elasticobjects\ create two or more JVM instances on the same slice obtained from
Mesos. Then, \elasticobjects\ instantiates the HyperDex on one additional Mesos
slice, and continues to monitor the performance of the HyperDex over the lifetime
of the elastic object. \elasticobjects\ may add additional nodes to HyperDex  
as necessary. \elasticobjects\ also enables administrators to be notified
if the utilization of the Mesos cluster exceeds or falls below (configurable) thresholds,
enabling the proactive addition of computing resources before 
the cluster runs out of slices.



\subsection{Load Balancing}\label{sec:loadbalancing}
Unlike websites or web services, where load balancing \emph{has to be performed}
on the server-side, \elasticobjects\ has the 
advantage that both client and server programs are pre-processed to 
generate stubs and skeletons respectively. Hence, we employ a \emph{hybrid}
load balancing model involving both stubs and skeletons -- note that \emph{all}
load balancing code is generated by the pre-processor, and the programmer \emph{does
not have to handle any aspect of it explicitly}. Please also note that this section
describes the simple load balancing techniques used in \elasticobjects, but we do not claim to have invented a new
load balancing algorithm.


On the server side, the runtime, while instantiating skeletons in an elastic object pool,
assigns monotonically increasing unique identifiers ($uid$) to each skeleton, and stores this information
in HyperDex. The skeleton with the lowest $uid$ is chosen by the runtime to be the leader of the \pool, 
called the \emph{sentinel}. This is similar to leader election algorithms that use a so-called 
``royal hierarchy'' among processes in a distributed system. The sentinel, in addition to performing all the regular functions (forwarding remote method invocations)
to \emph{its} object in the pool, also helps in load balancing.
The client stub created by the \elasticobjects\ preprocessor (see Section~\ref{sec:stubsskeletons})
 has the ability to communicate with the sentinel to invoke remote methods.
 While contacting the sentinel for the first time, the stub on the client JVM requests 
 the identities (IP address and port number) of the other skeletons in the pool from the sentinel. 
 
 For load-balancing on
 the client-side, the stub then re-directs 
 subsequent method invocations to other objects in the object pool either randomly or in
 a round-robin fashion. If an object has been 
 removed from the pool after its identity is sent to a stub, i.e., if the sending itself fails, the remote method invocation throws 
 an exception which is intercepted by the client stub. The stub then retries the invocation on 
 other objects including the sentinel. If all attempts to communicate with the elastic object pool
 fail, the exception is propagated to the client application.

For load-balancing on the server side, the sentinel is also responsible for collecting and periodically broadcasting the state of the pool -- number of objects, their identities and the number of pending invocations --  to the skeletons of all its members. We use the JGroups group communication system for broadcasts. If the sentinel notices that any skeleton is overloaded with respect to others, it instructs the skeleton to redirect
a portion of invocations to a set of other skeletons. To decide the number of invocations that have to be 
redirected from each overloaded skeleton, our implementation of the sentinel uses the first-fit greedy bin-packing approximation algorithm (See \url{http://en.wikipedia.org/wiki/Bin_packing_problem}). As mentioned in the previous paragraph, client-side load balancing occurs at the 
 stub while server-side load balancing involves skeletons and the sentinel which monitor the state of the JVM
 and that of the \pool\ to redirect incoming method invocations.


\subsection{Fault Tolerance}
Existing RMI
applications implement fault tolerance protocols on top of Java RMI's 
fault- and fault tolerance model, where objects typically reside in main memory,
and can crash in the middle of a remote method invocation. !e 
want to preserve it to make adoption of \elasticobjects\ easier. In short,
\elasticobjects\ \emph{does not hide/attempt to recover} from failures
of client objects, key-value store (HyperDex) or the server-side runtime processes
and propagates corresponding exceptions to the application. However, \elasticobjects\
\emph{attempts to recover} from failures of the sentinel and from Mesos-related failures. 
Sentinel failure triggers the leader election algorithm described 
in \ref{sec:loadbalancing} to elect a new sentinel, and mesos-related failures affect the 
addition/removal of new objects until Mesos recovers. 




%% file: evaluation.tex
\section{Evaluation}\label{sec:eval}

In this section, we evaluate the performance of \elasticobjects,
using metrics relevant to elasticity. Due to space limitations,
we refer the reader to our tech report for additional details~\cite{techreport}.

\subsection{Elasticity Metrics}

\emph{Measuring elasticity is different from measuring scalability}. (Recall that) Scalability is the ability of a 
distributed application to increase its ``performance'' proportionally (ideally linearly) 
with respect to the number of 
available resources, while elasticity is the ability of the application to adapt to 
increasing or decreasing workload; adding or removing resources to \emph{maintain} a specific level of ``performance'' or ``quality of service (QoS)''\cite{specmetrics}. Performance/QoS is specific to the application -- typically
a combination of throughput and latency. A highly elastic system can scale to 
include newer compute nodes, as well as quickly provision those nodes. There are no standard \emph{benchmarks}
for elasticity, but the Standard Performance Evaluation Corporation (SPEC)
has recommended elasticity \emph{metrics} for IaaS and PaaS clouds~\cite{specmetrics}.

\paragraph{Agility.}

This metric characterizes the ability of a 
system provisioned to be as close to the needs of the workload as possible~\cite{specmetrics}. 
Assuming a time interval $[t,t']$, which is divided into $N$
sub-intervals, Agility maintained over $[t,t']$ can be defined as:

\vspace{-8mm}
\[
\frac{1}{N}(\sum_{i=0}^{N}Excess(i) + \sum_{i=0}^{N}Shortage(i))
\]

\noindent where (1)$Excess(i)$ is the excess capacity for interval $i$ as determined by 
$Cap\_prov(i) - Req\_min(i)$, when $Cap\_prov(i) > Req\_min(i)$ 
and zero otherwise. (2) $Shortage(i)$ is the shortage capacity for interval 
$i$ as determined by 
$Req\_min(i) - Cap\_prov(i)$, when $Cap\_prov(i) < Req\_min(i)$ and zero otherwise. (3) $Req\_min(i)$ is the minimum capacity needed to 
meet an application's quality of service (QoS) at a given workload level for an 
interval $i$. (4) $Cap\_prov(i)$ is the recorded capacity provisioned for interval $i$, and 
(5) $N$ is the total number of data samples collected over a measurement period $[t,t']$, i.e., one sample
of both $Excess(i)$ and $Shortage(i)$ is collected per sub-interval of $[t, t']$.

Elasticity measures the shortage \emph{and} excess of computing
resources over a time period. For example, a value of elasticity of $2$ over $[t,t']$ when there is no 
excess means that there is a mean shortage of $2$ ``compute nodes'' over $[t,t']$. For an ideal system, agility  should be as close to zero as possible -- meaning that there is neither a shortage nor excess. 
Agility is a measurement of the ability to scale up and down while maintaining a 
specified QoS. The above definition of agility will not be valid in a context where the QoS 
is not met. It should be noted that there is ongoing debate over whether $Shortage$ and $Excess$ should be given equal 
weightage\cite{specmetrics} in the Agility metric, but there are disagreements over what the weights should be otherwise.


\paragraph{Provisioning Interval.}

Provisioning Interval is defined as the time needed to bring up or drop a resource. 
This is the time between initiating the request to bring up a new resource, and when the resource serves the first request. 

\begin{figure*}[htbp]
\subfloat[Pattern for abruptly changing workload for all four systems. The pattern remains the same, but the meaning and magnitude vary for the four systems.]{\includegraphics[width=0.48\textwidth]{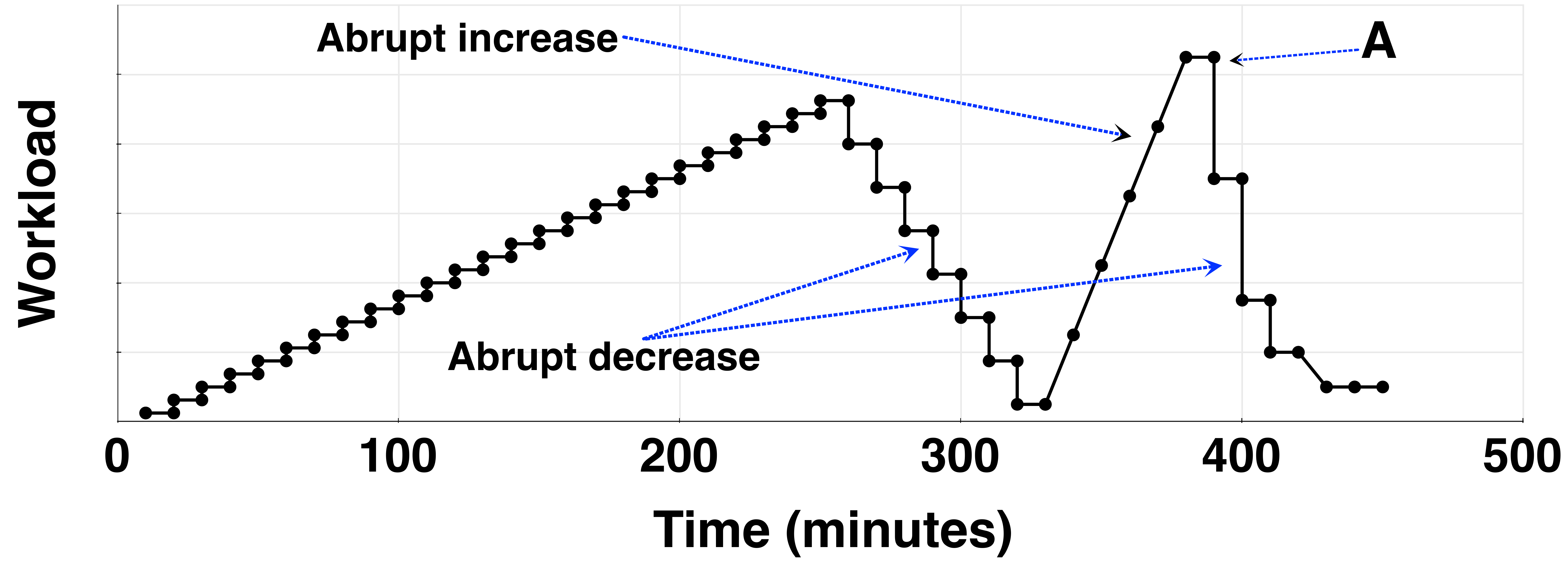}\label{fig:abrupt}}\hspace{2mm}
\subfloat[Cyclical workload example for all four systems. As in Figure~\ref{fig:abrupt}, the pattern remains the same for all four systems but the meaning and magnitude are different.]{\includegraphics[width=0.48\textwidth]{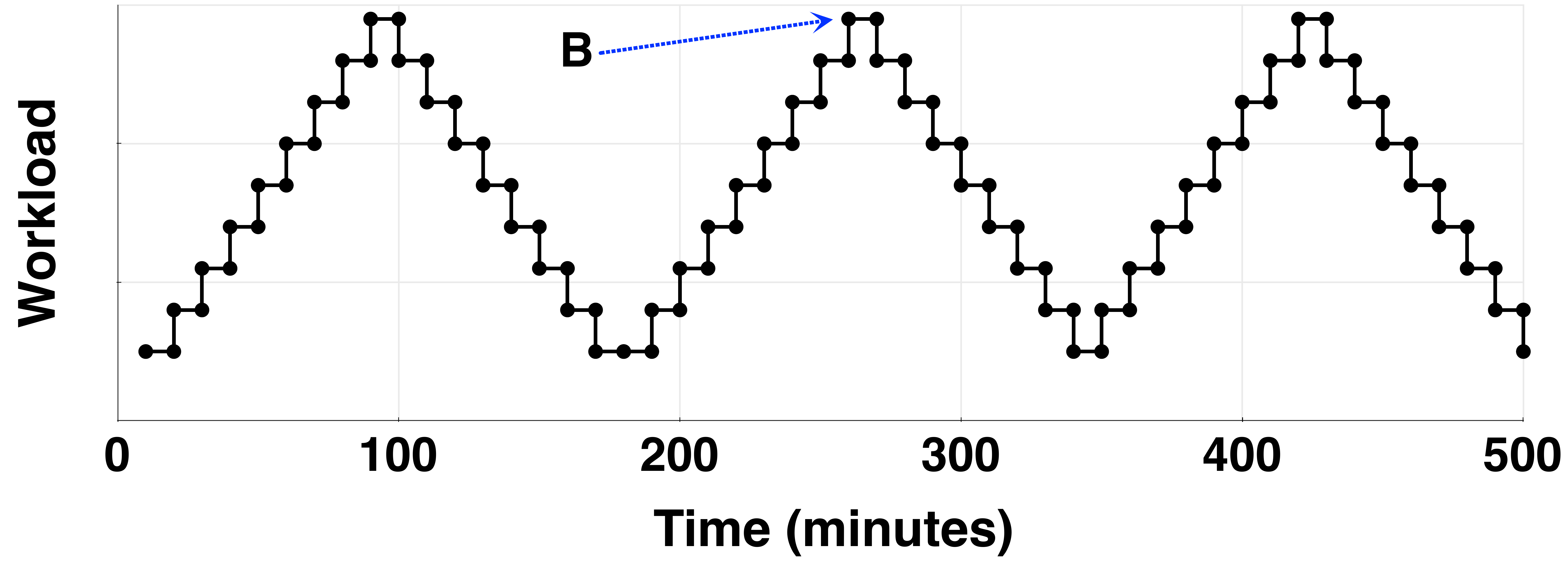}\label{fig:cyclic}} \\ \vspace{-3mm}
\subfloat[Marketcetera -- abrupt workload.]{\includegraphics[width=0.50\textwidth]{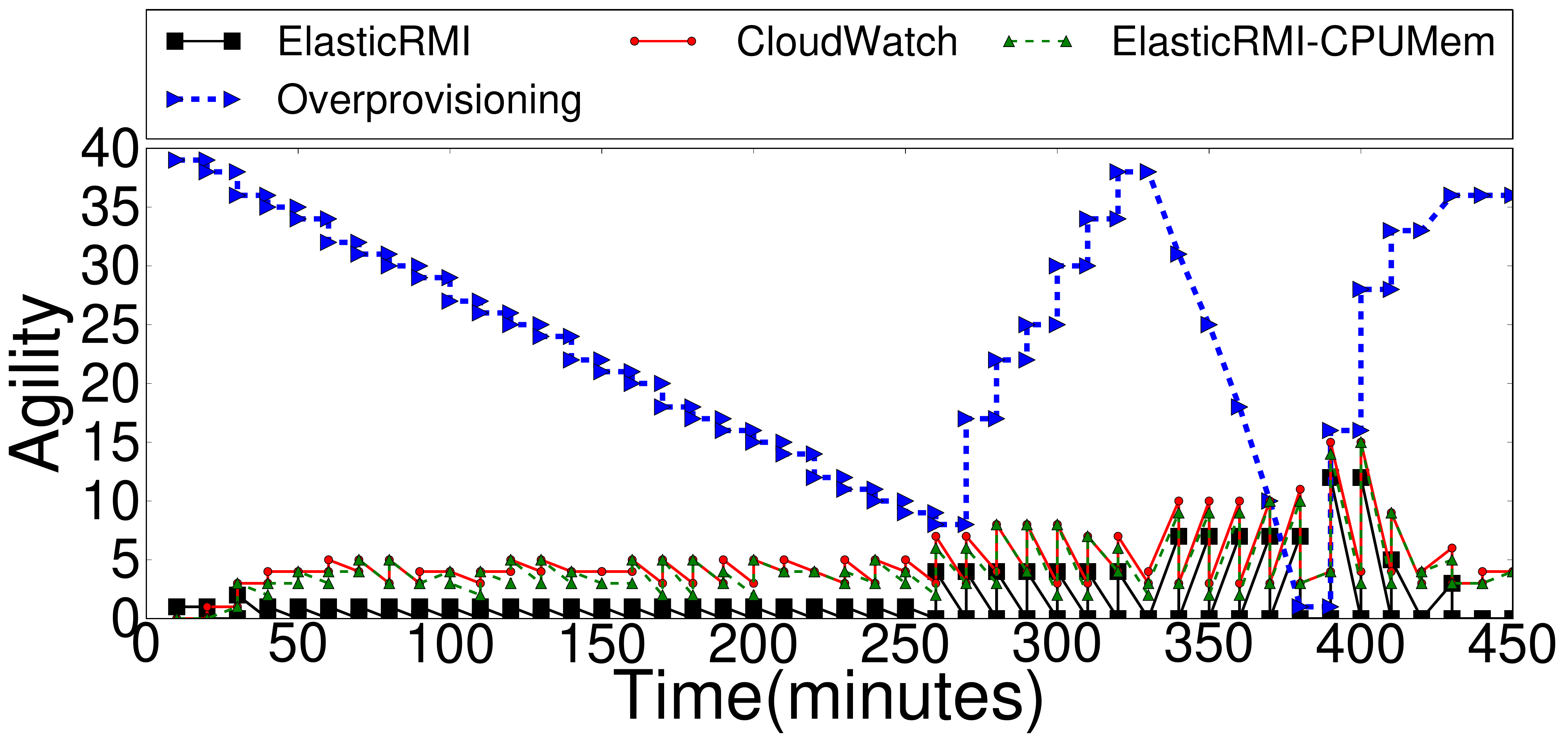}\label{fig:mk-abrupt-agility}}
\subfloat[Marketcetera -- cyclical workload.]{\includegraphics[width=0.50\textwidth]{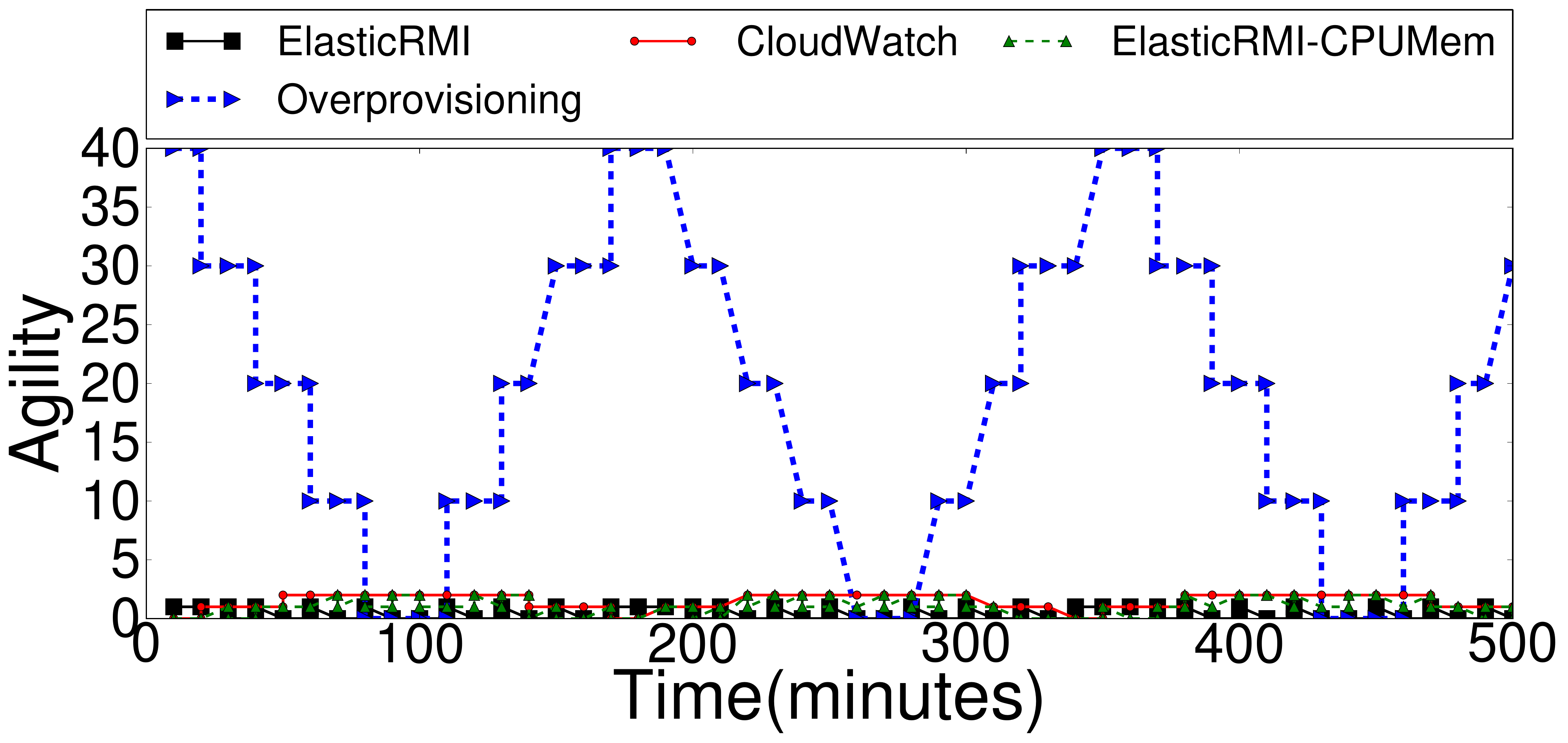}\label{fig:mk-cyclic-agility}} \\ \vspace{-3mm}
\subfloat[Hedwig -- abrupt workload.]{\includegraphics[width=0.50\textwidth]{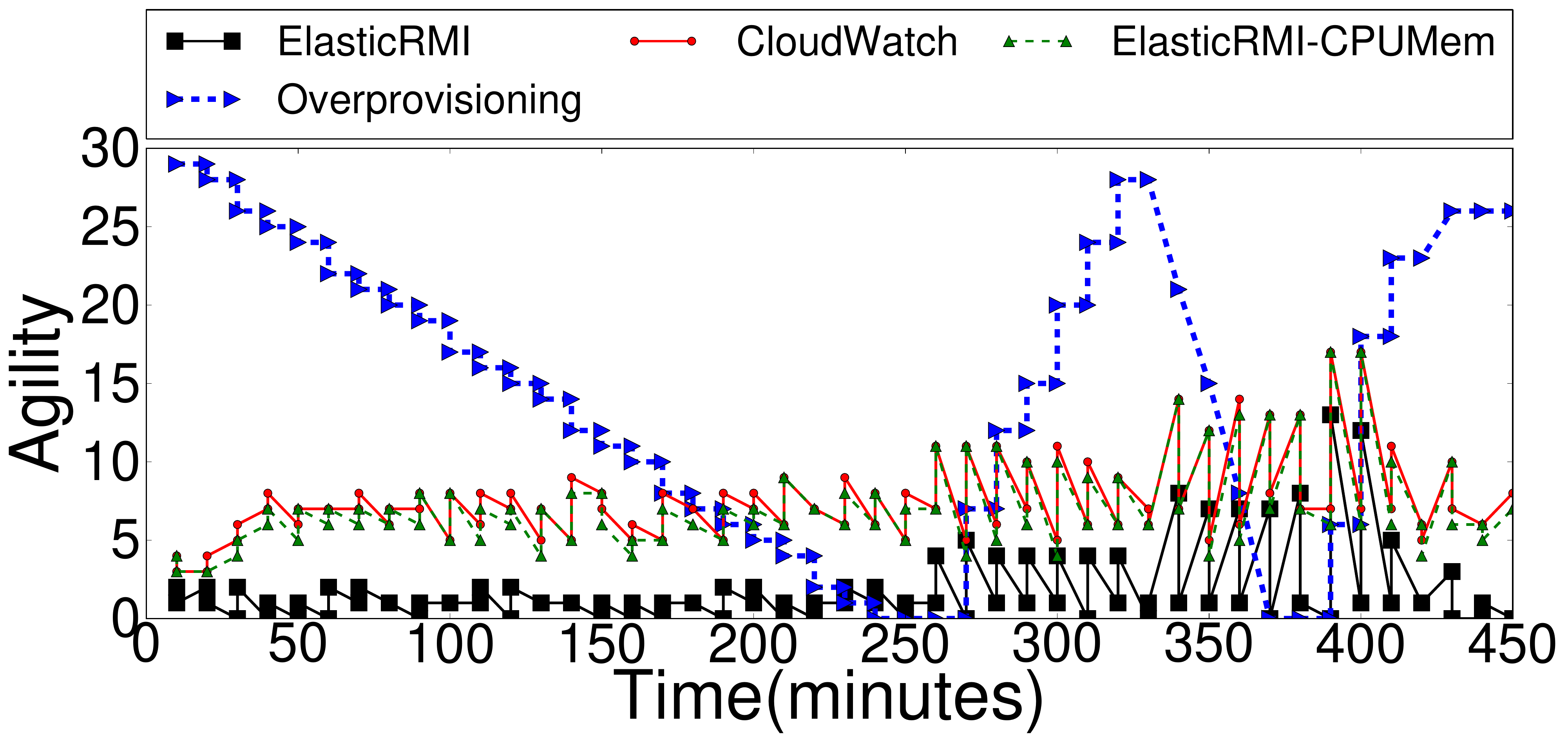}\label{fig:hedwigabruptagility}}
\subfloat[Hedwig -- cyclical workload.]{\includegraphics[width=0.50\textwidth]{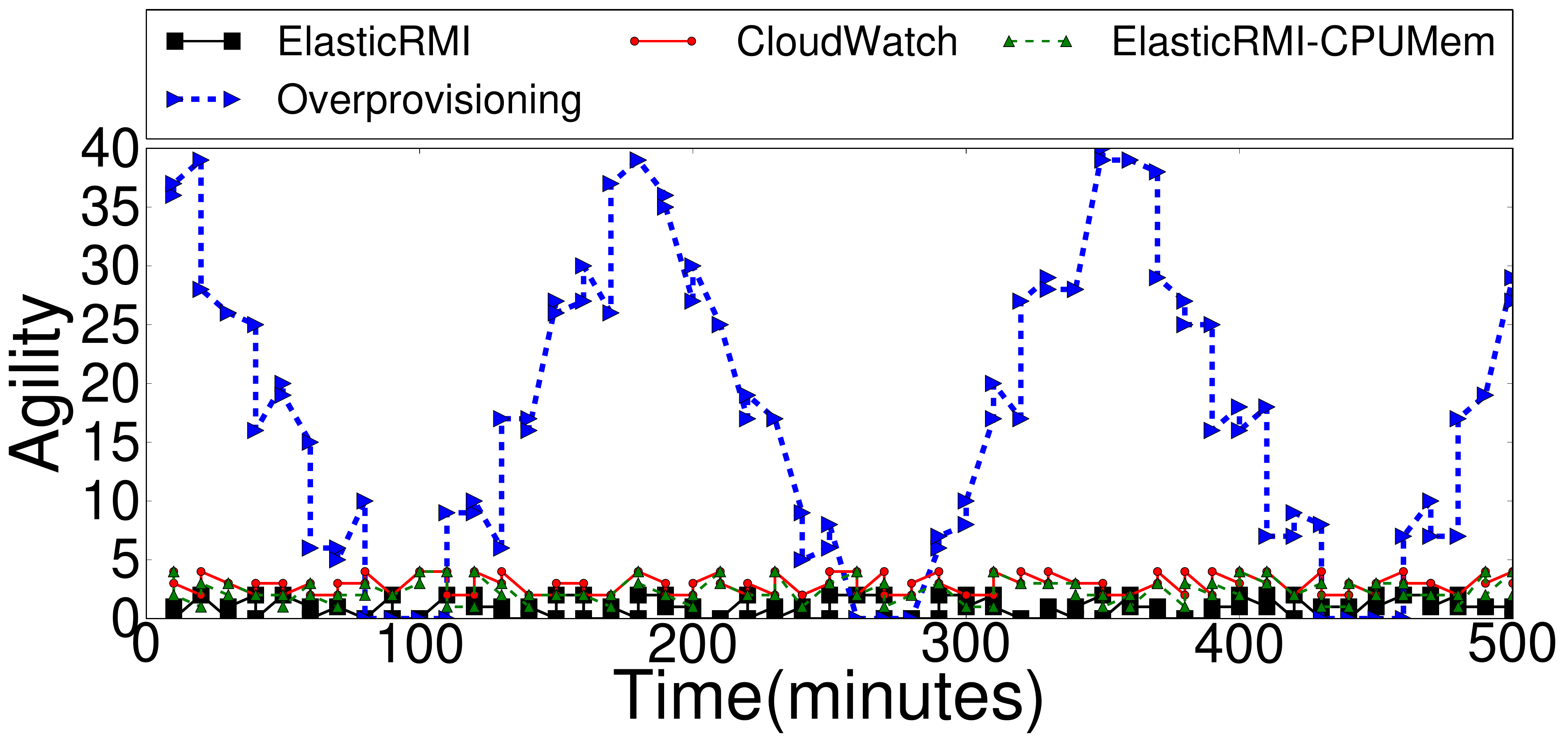}\label{fig:hedwigcyclicagility}} \\ \vspace{-3mm}
\subfloat[Paxos -- abrupt workload.]{\includegraphics[width=0.50\textwidth]{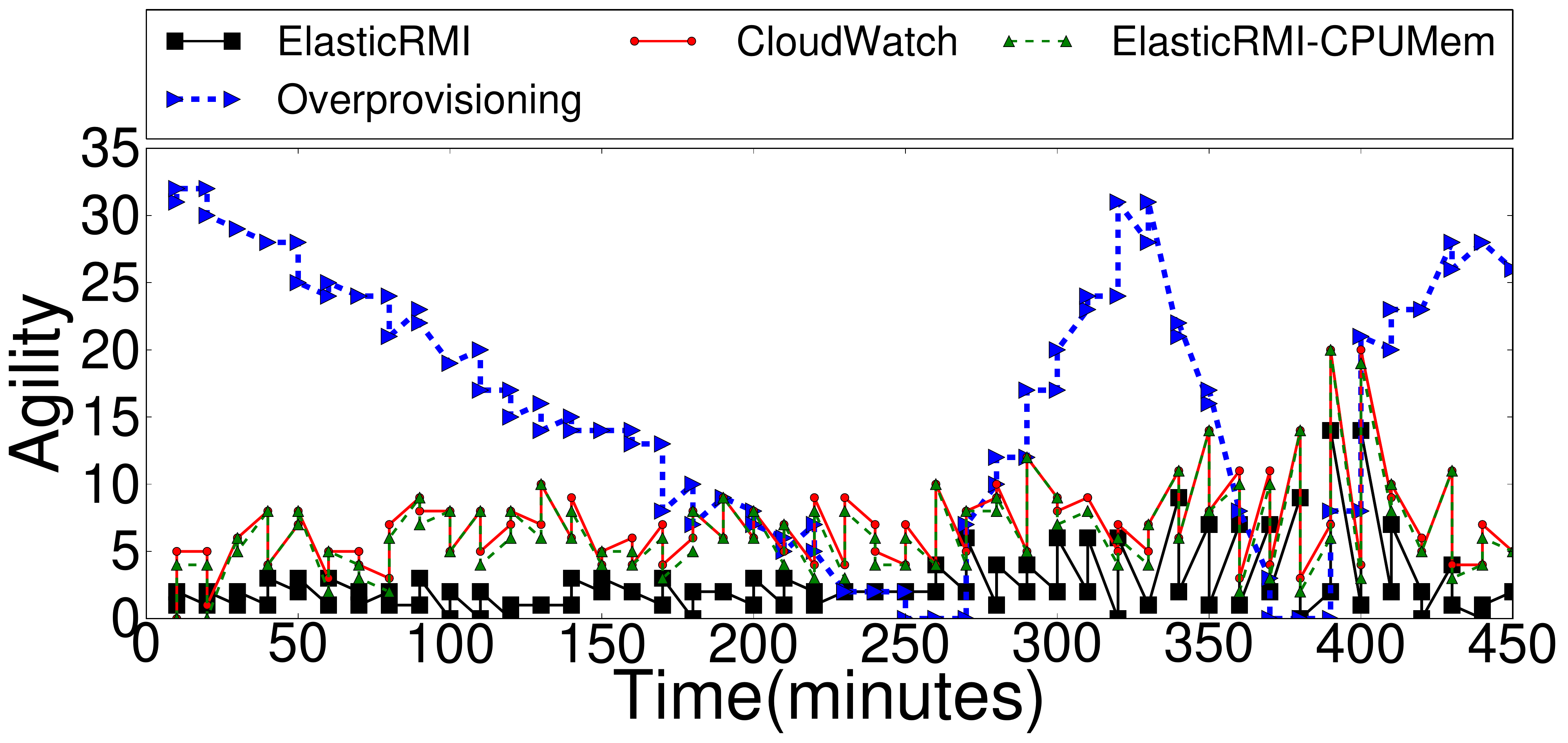}\label{fig:paxosabruptagility}}
\subfloat[Paxos -- cyclical workload.]{\includegraphics[width=0.50\textwidth]{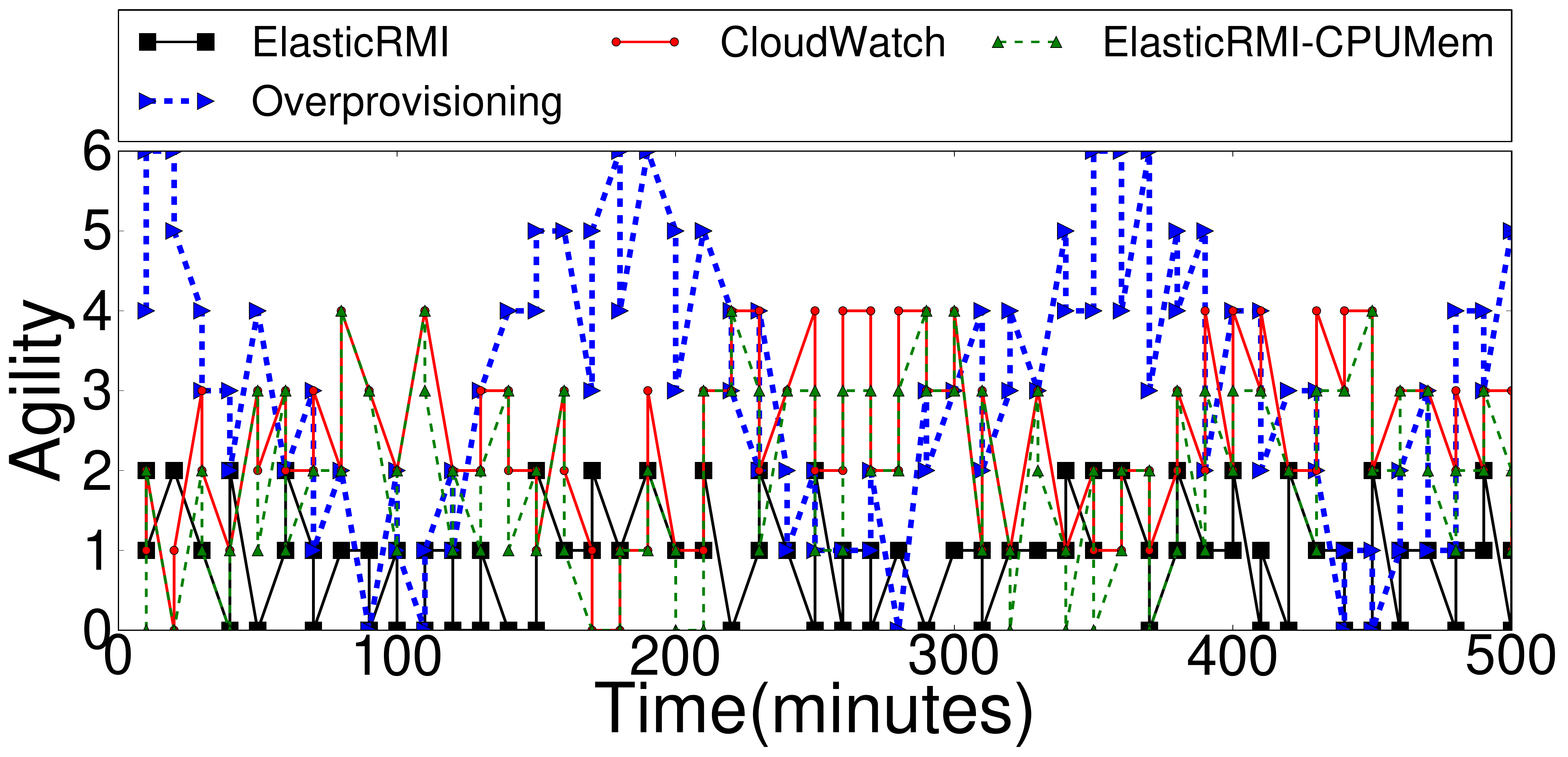}\label{fig:paxoscyclicagility}}\\ 
\subfloat[DCS -- abrupt workload.]{\includegraphics[width=0.50\textwidth]{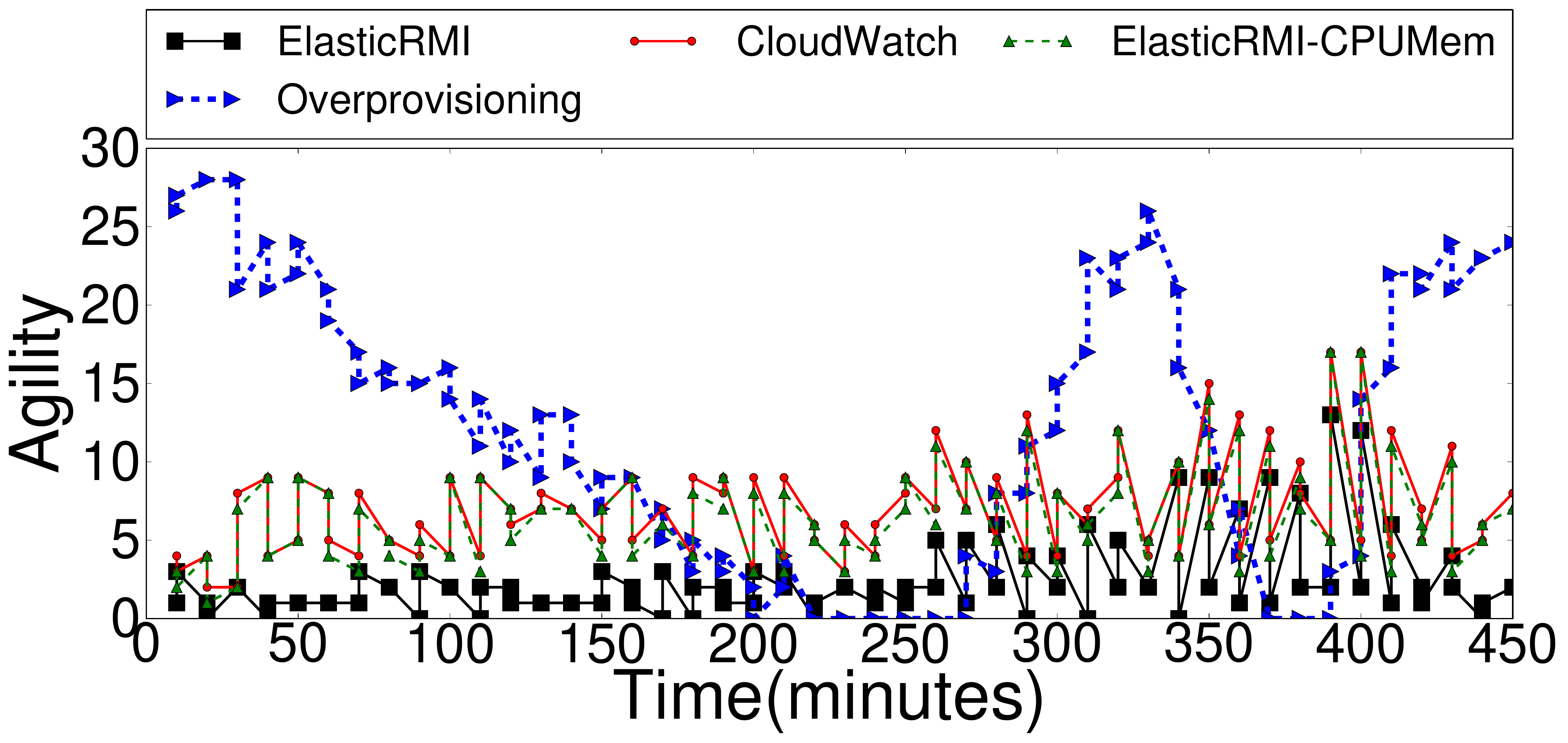}\label{fig:dcsabruptagility}}
\subfloat[DCS -- cyclical workload.]{\includegraphics[width=0.50\textwidth]{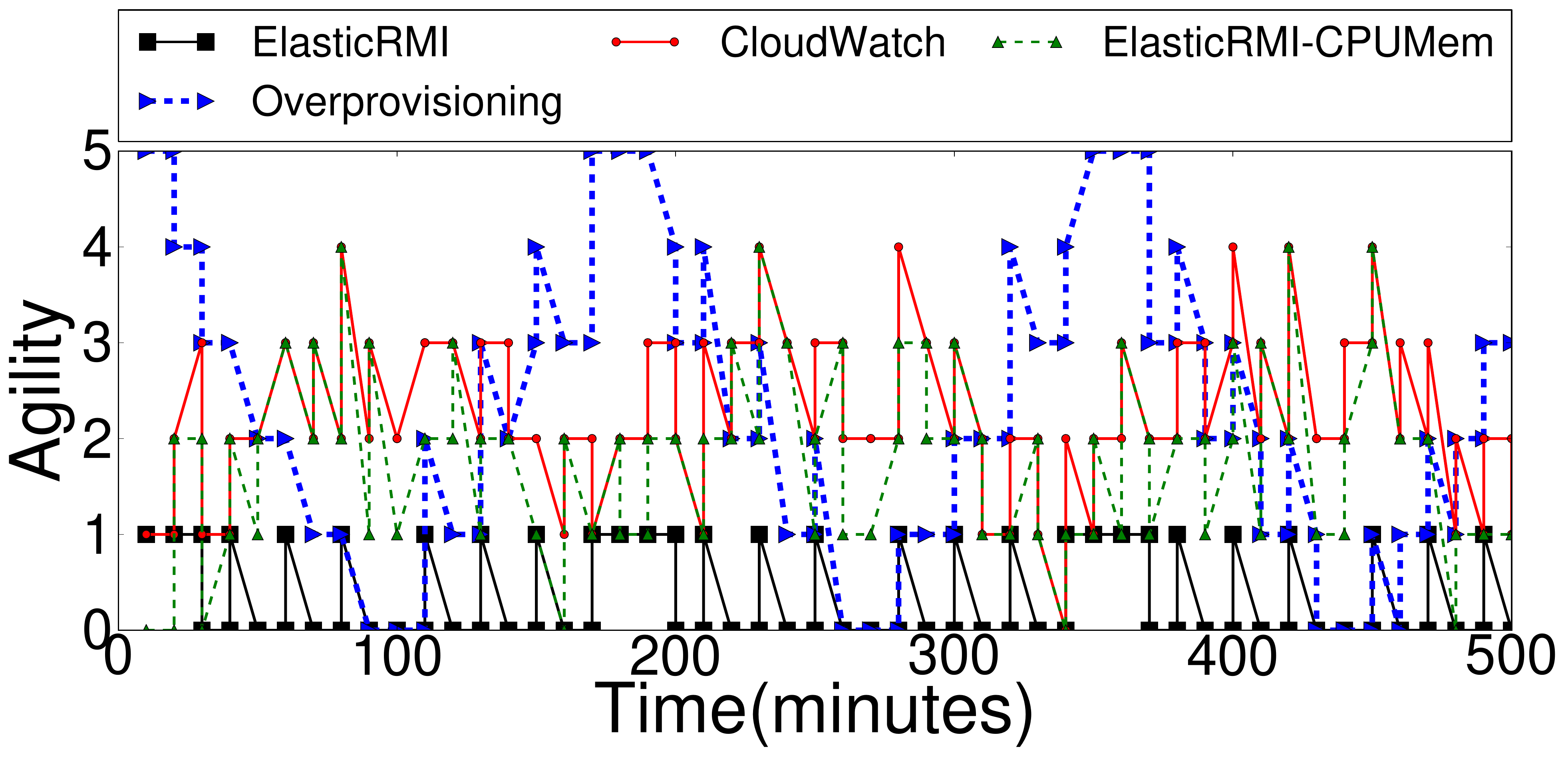}\label{fig:dcscyclicagility}}
\caption{Elasticity Benchmarking of Marketcetera order routing, Hedwig, Paxos and DCS. We compare the \elasticobjects\ implementation of these applications with two other systems described in Section~\ref{sec:compsys}.}\label{fig:allgraphs}
\end{figure*}


\subsection{\elasticobjects\ Applications for Evaluation and Workloads}\label{sec:apps}

We have re-implemented four \emph{existing} applications using \elasticobjects\ to add elasticity management components
to them. This does not involve altering the percentage of shared state or the frequency of accesses (reads or writes) to said state.


\paragraph{Marketcetera~\cite{marketcetera} Order Routing.} Marketcetera is an NYSE-recommended algorithmic trading platform. The order routing system is the component that accepts orders from traders/automated strategy engines and routes them to various markets (stock/commodity), brokers and other financial intermediaries. For fault-tolerance, the order is persisted (stored) on two nodes. The workload for this system is a set of trading orders generated by the simulator included in the community edition of Marketcetera~\cite{marketcetera}. 

\paragraph{Apache Hedwig~\cite{hedwig}.}
Hedwig is a topic-based publish-subscribe system designed for reliable and guaranteed at-most once 
delivery of messages from publishers to subscribers. Clients are associated with (publish to and subscribe from) a Hedwig instance (also referred to as a region), which consists of a number of servers called hubs. The hubs partition the topic ownership among themselves, and all publishes and subscribes to a topic must be done to its owning hub. The workload for
this system is a set of messages generated by the default Hedwig benchmark included in the implementation. 

\paragraph{Paxos~\cite{paxos}.}
Paxos is a family of protocols for solving consensus in a distributed system of unreliable processes. 
Consensus protocols are the basis for the state machine approach to distributed computing, and for our experiments
we implement Paxos using a widely-used specification by Kirsch and Amir~\cite{paxos}. The workload for this system is the default benchmark included
in \lstinline{libPaxos}~\cite{libpaxos}. 

\paragraph{DCS.} DCS is a distributed co-ordination service for datacenter applications, similar to Chubby~\cite{chubby} and Apache Zookeeper~\cite{zookeeper}.
DCS has a hierarchical name space which can be used for distributed configuration and synchronization. Updates are totally ordered. The workload 
for this system is the default benchmark included in Apache Zookeeper~\cite{zookeeper}.

\subsection{Workload Pattern}
To measure how
well the system adapts to the changing workload, we use two patterns shown in Figures~\ref{fig:abrupt}
and \ref{fig:cyclic}. These two patterns capture all common scenarios in elastic scaling which we have observed
by analyzing real world applications.
The abrupt pattern shown in Figure~\ref{fig:abrupt} has all possible scenarios regarding abrupt changes
in workload -- gradual non-cyclic increase, gradual decrease, rapid increases and rapid decrease in workload.
A cyclic change in workload is shown by the second pattern in Figure~\ref{fig:cyclic}. So, together
the patterns in Figures~\ref{fig:abrupt}
and \ref{fig:cyclic} exhaustively cover all elastic scaling scenarios we observed. Note however, that although the pattern 
remains the same for varying the workload while evaluating all the four systems, the magnitude differs depending on the benchmark used, i.e., 
the values of points A and B in Figures~\ref{fig:abrupt}
and \ref{fig:cyclic} are different for the four systems depending on the benchmark. Point A, for example, is 50,000 orders/s for Marketcetera, 75,000 updates/s for DCS,
24,000 consensus rounds/s for Paxos and 30,000 messages/s for Hedwig. We set Point B at 20\% above Point A -- note that the specific values of Points A and B are immaterial because we are only measuring adaptability and not peak performance.



\subsection{Overprovisioning and CloudWatch}\label{sec:compsys}

We compare the \elasticobjects\ implementation of the applications in 
Section~\ref{sec:apps} with the \emph{existing} implementations of the same
applications in two deployment scenarios -- (1) Overprovisioning and 
(2) Amazon AutoScaling + CloudWatch~\cite{autoscaling}\cite{cloudwatch}. The overprovisioning deployment scenario is 
similar to an ``oracle'' -- the \emph{peak} workload arrival rate i.e., point A for the abruptly changing workload
and point B for the cyclic workload are known a priori to the oracle; and the number of nodes required to meet
a desired QoS (throughput, latency) at A and B respectively is determined by the oracle through
experimental evaluation. The oracle then provisions the application on 
a fixed set of nodes -- the size of which is enough to maintain
the desired QoS even at the peak workload arrival rate (A and B respectively). In a nutshell,
the over provisioning scenario can be described as ``knowing future workload 
patterns and provisioning enough resources to meet its demands''.  Overprovisioning is 
the alternative to elastic scaling -- there are going to be excess provisioned resources when the workload is below
the peak (A and B), but provisioning latency is zero because all necessary resources are always provisioned. In the CloudWatch scenario, we use a monitoring
service -- Amazon CloudWatch to collect utilization metrics (CPU/Memory) from
the nodes in the cluster and use conditions on these metrics to decide whether to 
increase or decrease the number of nodes. The \elasticobjects\ implementation
of the above applications, however, uses a combination of resource utilization and 
application-level properties specific to Marketcetera, DCS, Paxos and Hedwig respectively to 
decide on elastic scaling. Since \elasticobjects\ and CloudWatch are two different systems,
we also compare the \elasticobjects\ implementation of the four applications with another version,
referred to as ElasticRMI-CPUMem in Figure~\ref{fig:allgraphs}, 
where no application-level properties are used but only the the conditions based on CPU/Memory utilization 
in CloudWatch are used. 

\begin{figure*}[htb]
\subfloat[Provisioning latency -- Abrupt Workload]
{\includegraphics[width=0.49\textwidth]{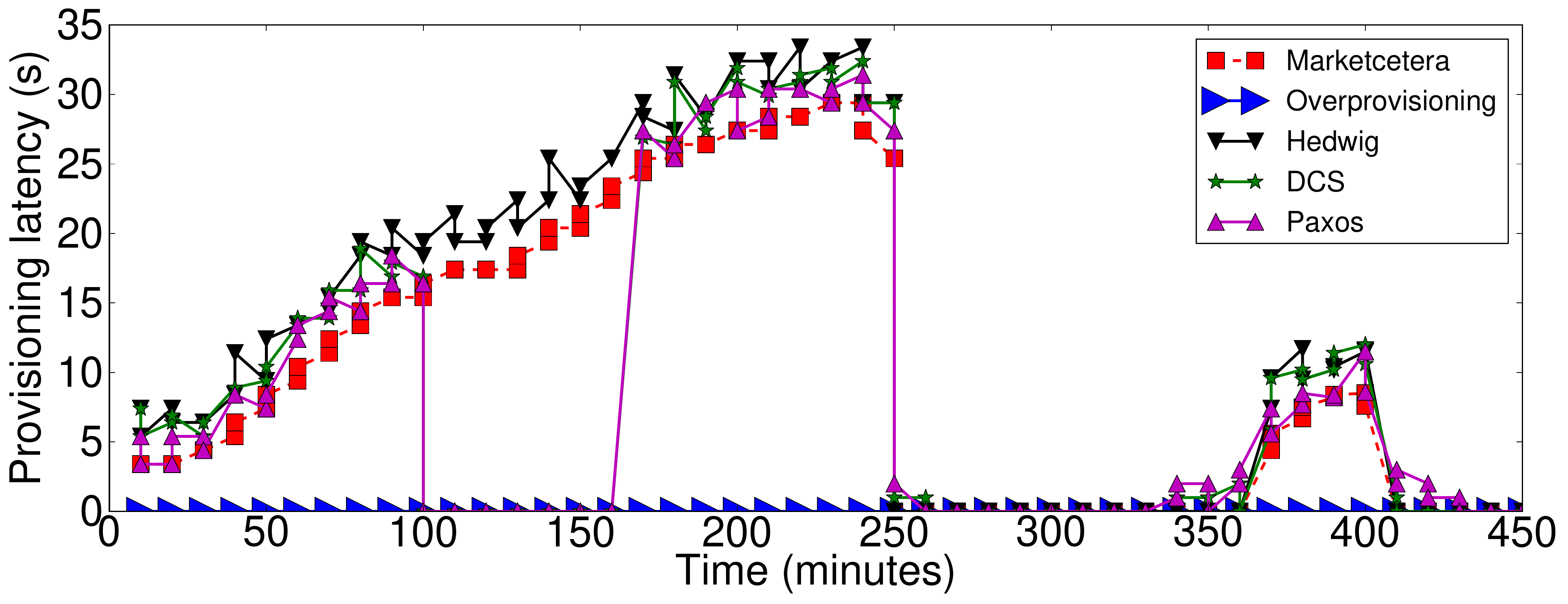}\label{fig:provabr}}
\subfloat[Provisioning latency -- Cyclic Workload]
{\includegraphics[width=0.49\textwidth]{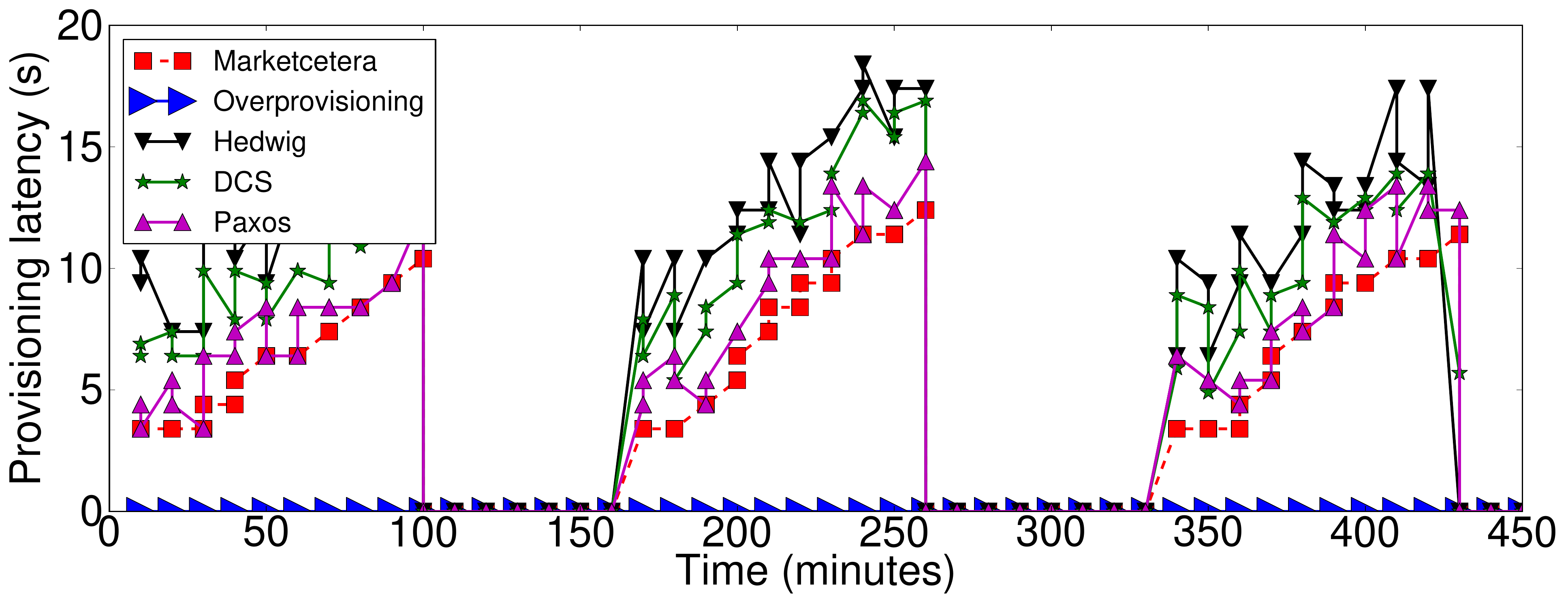}\label{fig:provcyc}} 
\caption{Provisioning latency in seconds for ElasticRMI and Overprovisioning (which is always 0). Provisioning 
latency for Amazon CloudWatch is not plotted because it is in several minutes and hence well 
above that of both ElasticRMI and Overprovisioning. You can see repeating patterns corresponding to
the cyclic workload.}\label{fig:provlat}
\end{figure*}

\subsection{Agility Results}
In this section, we compare the Agility of the ElasticRMI implementation of all four systems
against Overprovisioning, CloudWatch and ElasticRMI-CPUMem.

\paragraph{Marketcetera Order Routing.}

The relevant QoS metrics for the order routing subsystem are order routing throughput,
which is the number of orders routed from traders to brokers/exchanges per second and 
order propagation latency, which is the time taken for an order to propagate from the
sender to the receiver. We compare the elasticity of the three deployments of the order routing system described 
in Section~\ref{sec:compsys}. The results are illustrated in Figure~\ref{fig:allgraphs}.
Figures~\ref{fig:mk-abrupt-agility} and \ref{fig:mk-cyclic-agility} plot
the agility over the same time period as in Figures~\ref{fig:abrupt} and \ref{fig:cyclic} for 
all the four deployments. From Figures~\ref{fig:mk-abrupt-agility} and \ref{fig:mk-cyclic-agility},
we observe that the agility of \elasticobjects\ is better than CloudWatch, ElasticRMI-CPUMem and 
overprovisioning. Ideally, agility must be zero, because agility is essentially a 
combination of resource wastage or resource under-provisioning. We observe that for
abruptly changing workloads, agility of \elasticobjects\ is close to 1 most of the time, 
and increases to 5 during abrupt changes in workload. We also observe that the the 
agility of \elasticobjects\ oscillates between 0 and a positive value frequently.
This proves that the elastic scaling mechanisms of \elasticobjects\ perform 
well in trying to achieve optimal resource utilization, i.e., react aggressively by trying to
 push agility to zero. In summary, the average agility of \elasticobjects\ for abruptly changing workload is 1.37. As expected, the agility of overprovisioning is the worst, up to ~24$\times$ that of \elasticobjects. This is not 
 surprising, because its agility does reach zero at peak workload, when the 
 agility of \elasticobjects\ is 5, thus illustrating that overprovisioning optimizes
 for peak workloads. The average agility of overprovisioning is 17.2 for the cyclical workload and 24.1 for abruptly changing workload. CloudWatch performs much better than overprovisioning, but it is less agile 
 than \elasticobjects. Its agility is approximately 3.4$\times$ that of \elasticobjects\ on
 average for abruptly changing workloads, and it does not oscillate to zero frequently like \elasticobjects. The 
 agility of ElasticRMI-CPUMem is approximately equal to CloudWatch and is 3.02$\times$ that of
 \elasticobjects\ on average -- this is in spite of ElasticRMI-CPUMem and CloudWatch being different systems.
 having different provisioning latencies. This is because the same conditions are used to decide
 on elastic scaling and because the provisioning latency of CloudWatch is well within the sampling interval of 10 minutes used in Figure~\ref{fig:allgraphs}.
 
 Figure~\ref{fig:mk-cyclic-agility} shows that the agility of \elasticobjects\ is again better
 than that of CloudWatch and overprovisioning for cyclic workloads. We also observe that 
 as in the case of abrupt workloads, the agility of \elasticobjects\ tends to decrease to
 zero more frequently than the other two deployments. Figure~\ref{fig:mk-cyclic-agility}
 also demonstrates the oscillating pattern in the agility of overprovisioning -- 
 the initial agility is high (and comes from $Excess$), and as the 
 workload increases, $Excess$ decreases, thereby decreasing Agility and bringing it
 to zero corresponding to Point B. Then $Excess$ increases again as the workload
 decreases thereby increasing Agility. This repeats three times. As expected,
 the agility of ElasticRMI-CPUMem is similar to that of CloudWatch.
 
\paragraph{Hedwig.}

The relevant QoS metrics for Hedwig are also throughput and latency -- the number of 
messages published per second and time taken for the message to propagate from the 
publisher to the subscriber. Figures~\ref{fig:hedwigabruptagility} and 
\ref{fig:hedwigcyclicagility} illustrate the agility corresponding to our experiments
with Hedwig. From Figures~\ref{fig:hedwigabruptagility} and 
\ref{fig:hedwigcyclicagility}, we observe similar trends as in the case of Marketcetera order
processing. \elasticobjects\ has lower agility values than the other two deployments, and the agility of 
\elasticobjects\ tends to oscillate between zero and a positive value. The agility values of
CloudWatch are more than 4.5$\times$ that of \elasticobjects, on average for abrupt workloads
and 3$\times$ that of \elasticobjects\ for cyclic workloads. As expected, the agility
of over provisioning is the highest, and is worse than the values observed for Marketcetera
in the case of cyclic workloads. We also observe a similar oscillating trend in the agility values
of the overprovisioning deployment as in Marketcetera, but the agility values oscillate more
frequently because $Req\_min(i)$ -- the minimum capacity needed to maintain QoS under a certain 
workload changes more erratically than Marketcetera due to the replication and at-most once
guarantees provided by Hedwig for delivered messages.

\paragraph{Paxos}

The relevant QoS metrics for Paxos are the number of consensus rounds 
executed successfully per second, and the time taken to execute a consensus round.
Figures~\ref{fig:paxosabruptagility} and 
\ref{fig:paxoscyclicagility} illustrate the agility corresponding to our experiments
with Paxos. From Figures~\ref{fig:paxosabruptagility} and 
\ref{fig:paxoscyclicagility}, we observe similar trends to Hedwig and Marketcetera.
The agility of CloudWatch in this case is ~6.6$\times$ than of \elasticobjects, on average for abrupt workloads
and ~2.2$\times$ that of \elasticobjects\ for cyclic workloads. We also observe that the agility of
\elasticobjects\ returns to zero (the ideal agility) most frequently among the three deployments.

\paragraph{DCS}

The relevant QoS metrics for DCS are the number of updates to the hierarchical
name-space per second and the end-to-end latency to perform an update as measured
from the client. Figures~\ref{fig:dcsabruptagility} and 
\ref{fig:dcscyclicagility} illustrate the agility corresponding to our experiments
with DCS. From Figures~\ref{fig:dcsabruptagility} and 
\ref{fig:dcscyclicagility}, we observe that the agility of CloudWatch in this case is ~7.2$\times$ than of \elasticobjects, on average for abrupt workloads
and ~3.2$\times$ that of \elasticobjects\ for cyclic workloads. 




\subsection{Provisioning Latency}
Figures~\ref{fig:provabr} and \ref{fig:provcyc} plot the provisioning latency of \elasticobjects
 for both abrupt and cyclic workloads. We observe that the provisioning latency of
 \elasticobjects\ is less than 30 seconds in all cases, which compares very favorably to
 the time needed to provision new VM instances in Amazon CloudWatch (which is in the order of
 several minutes, and hence omitted from Figure~\ref{fig:provlat}). Provisioning latency is zero for the overprovisioning scenario, and that is the main purpose of overprovisioning -- to have resources
 always ready and available. Also, we observe that as the 
 workload increases, provisioning interval also increases, due to the overhead in determining the 
 remote method calls that have to be redirected and also due to increasing demands on the 
 resources of the sentinel object in \elasticobjects's object pools.



%% file: relatedwork.tex
\section{Related Work}\label{sec:related}

J-Orchestra~\cite{jorchestra,dthreads} automatically partitions Java applications and makes them into distributed applications
running on distinct JVMs, by using byte code transformations to change local method calls into
distributed method calls as necessary. The key distinctions between J-Orchestra and \elasticobjects\
are that (1) J-Orchestra tackles the complex problem of
automatic distribution of Java programs while \elasticobjects\ aims to add elasticity to already
distributed programs and (2) \elasticobjects\ partitions 
different invocations of a single remote method.

 Self-Replicating Objects (SROs)~\cite{sro-ecoop10} is a new
elastic concurrent programming abstraction. An SRO is similar to an ordinary .NET object exposing
an arbitrary API but exploits multicore CPUs by automatically partitioning its
state into a set of replicas that can handle method calls (local and remote) in parallel,
and merging replicas before processing calls that cannot execute in replicated state.
SRO also does not require developers to explicitly protect access to shared data; the 
runtime makes all the decisions on synchronization, scheduling and splitting/merging state. Live Distributed Objects (LDO)~\cite{ldo-ecoop08} is a new programming paradigm and a 
platform, in which instances of 
distributed protocols are modeled as Òlive distributed objectsÓ. Live objects can be used
to represent distributed multi-party protocols and application components. Shared-state 
and synchronization between the objects is maintained using Quicksilver~\cite{quicksilver}, a group
communication system. Automatic scaling is not supported and must be explicitly implemented
by the programmer using the abstractions provided by LDO.

Quality Objects (QuO)~\cite{quo} is a seminal framework for providing quality of service (QoS) in network-centric distributed applications.
When the requirements are not being met, QuO provides the ability to adapt at many levels in the system,
including the middleware layer responsible for message transmission. In contrast to QuO, \elasticobjects\ attempts to increase quality of service by 
changing the size of the remote object pool, and does not change the protocols used to
transmit remote method invocations.

%% file: conclusion.tex
\section{Conclusions}\label{sec:conclusions}

We have described the design and implementation of \elasticobjects\
and have demonstrated through empirical evaluation using real-world
applications that it is effective in engineering elastic distributed 
applications. Our empirical evaluation also demonstrates that relying 
solely on externally observable metrics like CPU/RAM/network utilization 
decreases elasticity, as demonstrated by the high agility values of CloudWatch.
We have shown that our implementation of \elasticobjects\
reduces resource wastage, and is sufficiently agile to meet 
the demands of applications with dynamically varying workloads.
Through an implementation using Apache Mesos, we ensure portability of 
\elasticobjects\ applications across Mesos installations, whether 
it is a private datacenter or a public cloud or a hybrid deployment 
between private data centers and public clouds. We have demonstrated that
\elasticobjects\ applications can use fine-grained application specific
metrics without revealing those metrics to the cloud infrastructure provider,
unlike CloudWatch.

%% file: elasticrmi.bbl
\begin{thebibliography}{10}

\bibitem{cloudwatch}
{Amazon Web Services (AWS) Inc.}
\newblock {Amazon CloudWatch}.
\newblock {\em {\url{http://aws.amazon.com/cloudwatch/}}}, 2012.

\bibitem{pig}
{Apache Pig}.
\newblock {\url{http://pig.apache.org}}.
\newblock 2013.

\bibitem{thrift}
{Apache Thrift}.
\newblock {\url{http://thrift.apache.org/}}.
\newblock 2012.

\bibitem{autoscaling}
{AWS Inc.}
\newblock {Amazon Auto Scaling}.
\newblock {\em {\url{http://aws.amazon.com/autoscaling/}}}, 2012.

\bibitem{mesos}
{B.~Hindman and A.~Konwinski and M.~Zaharia and A.~Ghodsi and A.~Joseph and
  R.~Katz and S.~Shenker and I.~Stoica}.
\newblock {Mesos: A Platform for Fine-grained Resource Sharing in the Data
  Center}.
\newblock In {\em NSDI 2011. \url{http://incubator.apache.org/mesos/}}.

\bibitem{quo}
{BBN Technologies}.
\newblock {Quality Objects (QuO)}.
\newblock {\em {\url{http://quo.bbn.com/}}}, 2006.

\bibitem{chubby}
M.~Burrows.
\newblock {The Chubby Lock Service for Loosely-coupled Distributed Systems}.
\newblock In {\em OSDI 2006}.

\bibitem{jorchestra}
{E.~Tilevich and Y.~Smaragdakis}.
\newblock {J-Orchestra: Automatic Java Application Partitioning}.
\newblock In {\em {ECOOP 2002}}.

\bibitem{dthreads}
{E. Tilevich and Y.~Smaragdakis}.
\newblock {Portable and EfÞcient Distributed Threads for Java}.
\newblock In {\em {MIDDLEWARE 2004}}.

\bibitem{emr}
{Elastic Map Reduce}.
\newblock {\url{http://aws.amazon.com/elasticmapreduce/}}.
\newblock 2013.

\bibitem{marketcetera}
{G.~Miller and T.~Kuznets and R.~Agostino}.
\newblock {Marketcetera Automated Trading Platform}.
\newblock {\em {\url{http://www.marketcetera.com/site/}}}, 2012.

\bibitem{giraph}
{Giraph}.
\newblock {\url{http://incubator.apache.org/giraph/}}.
\newblock 2013.

\bibitem{hadoop}
{Hadoop}.
\newblock {\url{http://hadoop.apache.org}}.
\newblock 2013.

\bibitem{hedwig}
{Hedwig}.
\newblock \url{https://cwiki.apache.org/ZOOKEEPER/hedwig.html}.
\newblock 2013.

\bibitem{paxos}
{J.~Kirsch and Y.~Amir}.
\newblock {Paxos for Systems Builders}.
\newblock {\em \url{http://www.cnds.jhu.edu/pub/papers/cnds-2008-2.pdf}}, 2008.

\bibitem{ramcloud}
{J.Ousterhout et. al.}
\newblock {The Case for RAMCloud}.
\newblock {\em {CACM}}, 2011.

\bibitem{sro-ecoop10}
{K.~Ostrowski and C.~Sakoda and K.~Birman}.
\newblock {Self-replicating Objects for Multicore Platforms}.
\newblock In {\em ECOOP 2010}.

\bibitem{quicksilver}
{K.~Ostrowski and K.~Birman and D.~Dolev}.
\newblock {Quicksilver Scalable Multicast (QSM)}.
\newblock In {\em NCA 2008}.

\bibitem{ldo-ecoop08}
{K.~Ostrowski and K.~Birman and D.~Dolev and J.~Ahnn}.
\newblock {Programming with Live Distributed Objects}.
\newblock In {\em ECOOP 2008}.

\bibitem{techreport}
{K. R. Jayaram}.
\newblock {Elastic Remote Methods}.
\newblock {\em {Technical Report. Available from
  \url{http://www.jayaramkr.com/elasticrmi}}}, 2013.

\bibitem{libpaxos}
{LibPaxos}.
\newblock {\url{http://libpaxos.sourceforge.net/}}.
\newblock 2013.

\bibitem{memcached}
{Memcached}.
\newblock {\url{http://www.memcached.org}}.
\newblock 2013.

\bibitem{kvstore}
{R.~Escriva and B.~Wong and E.~Sirer}.
\newblock {HyperDex: a Distributed, Searchable Key-Value Store}.
\newblock In {\em SIGCOMM 2012}.

\bibitem{specmetrics}
{SPEC Open Systems Group (OSG)}.
\newblock {Report on Cloud Computing to the OSG Steering Committee}.
\newblock {\em
  {\url{http://www.spec.org/osgcloud/docs/osgcloudwgreport20120410.pdf}}},
  2012.

\bibitem{zookeeper}
Zookeeper.
\newblock \url{http://zookeeper.apache.org/}.
\newblock 2013.

\end{thebibliography}
